\documentclass[11pt]{article}
\usepackage{moriond,epsfig}

\def\be{\begin{equation}}
\def\ee{\end{equation}}
\def\bea{\begin{eqnarray}}
\def\eea{\end{eqnarray}}

\begin{document}
\vspace*{4cm}
\title{The First RHIC Spin Run--at PHENIX and PP2PP}

\author{Gerry Bunce\footnote{bunce@bnl.gov}}

\address{Physics Department and RIKEN BNL Research Center, Brookhaven National Laboratory\\
Upton, NY 11973-5000, USA}

\maketitle\abstracts{
The RHIC spin program will probe the spin structure of
the proton with polarized quarks and gluons by colliding
beams of polarized protons at Brookhaven.  The first
collisions, at root(s)=200 GeV, were recorded this year
in December 2001 and January 2002.  This report
describes the first run, and our plans for the next five+ years,
with planned runs at root(s)=200 and 500 GeV, and luminosity
L=2 x 10$^{32}~cm^{-2}s^{-1}$.
See also the presentation at Moriond by Bernd Surrow on the first run
with the STAR detector.  Here we include a first look at
measurements with the PHENIX and PP2PP detectors.
}

\section{Introduction}
The spin program of the Relativistic Heavy Ion Collider (RHIC) 
at BNL collides polarized proton beams from $\sqrt{s}=50$\,GeV to
$500$\,GeV, with expected luminosity of $L=2 \times 10^{32}$\,cm$^{-2}$\,s$^{-1}$
for the top energy collisions.  At these energies and luminosity,
the collisions at high transverse momentum can be described as
collisions of polarized partons, quarks and gluons, where the
polarized quarks and gluons of the protons in one beam probe
the spin structure of the protons in the other beam.  RHIC spin
provides a new laboratory to study the proton, compared to earlier and ongoing
extensive studies of nucleon spin structure using polarized lepton
beams incident on polarized proton and neutron targets.  The earlier lepton studies
gave us the major surprise that on average only about 25\% of the
proton spin is carried by the quarks and antiquarks in the proton.
Since the lepton experiments probe the nucleon via a virtual photon,
these {\it deep inelastic} lepton
experiments are sensitive to the quark distributions through electric
charge, and are not directly sensitive to the gluon contribution to
the proton spin.  With the strongly interacting quark and gluon probes at RHIC,
a major focus of our program is to measure the gluon polarization through
the {\it deep inelastic} scattering of polarized quarks from polarized protons.
Reference \cite{bunce} previews the RHIC spin program.

The cross section for the collision of high energy polarized protons,
producing particles at large $p_T$, factorizes similarly to the 
perturbative QCD description
of unpolarized proton-antiproton collisions at the Tevatron, for example.  
When an asymmetry is formed by comparing
the cross section for particles produced for 
collisions of protons with the same helicities,
with the cross section for particles produced for differing beam helicities,
we can directly measure the proton spin structure.
     
We will now describe the first RHIC spin run.

\section{Accelerating and Measuring Polarization}

\begin{figure}[ht]
\setlength{\unitlength}{\textwidth}
\begin{picture} (0.7,0.30)
\put (0.025,0.025){\mbox{\epsfig{figure=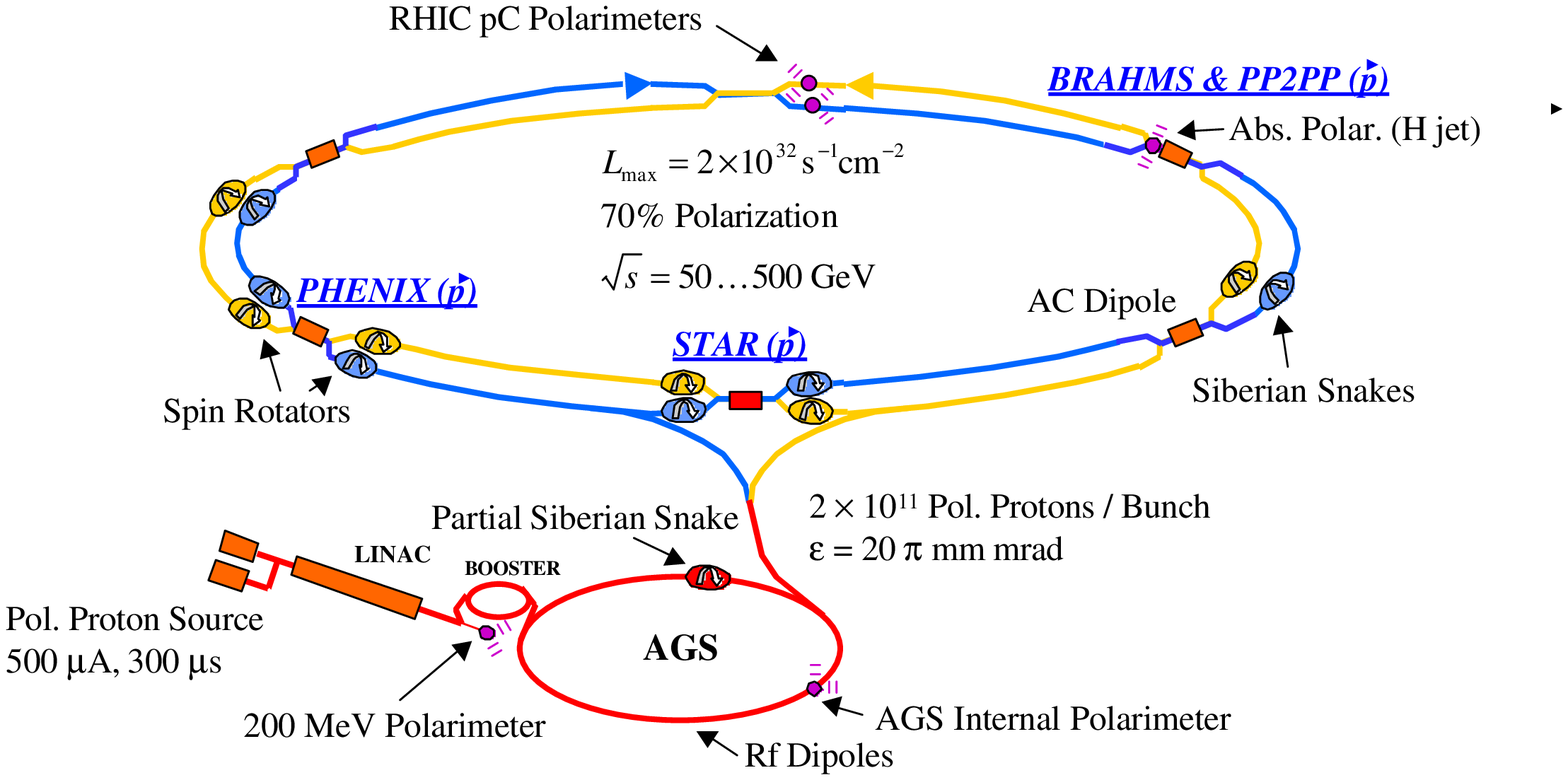,width=8.0cm,clip=}}}
\put (0.575,-0.015){\mbox{\epsfig{figure=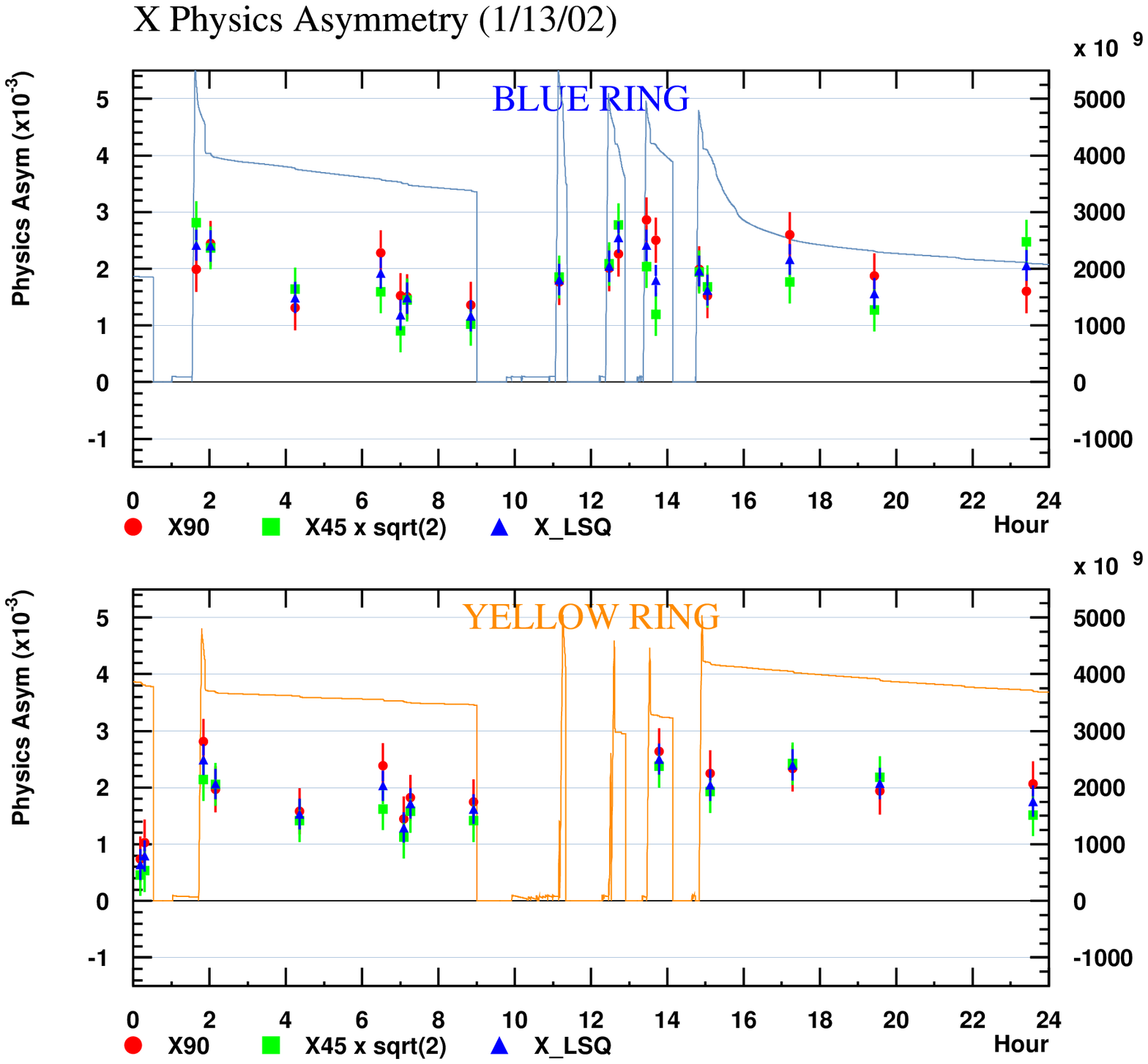,width=6.0cm,clip=}}}
\end{picture}
\caption{\it Schematic of the RHIC spin accelerator complex (top) and measurements of the
polarization with the beam intensity overlaid (bottom).  For the figure on the right, the
solid lines record the intensity in RHIC, referring to the axis on the right side of the
figure.  Independent polarimeter measurements of the asymmetry are shown as data points,
referring to the axis on the left side of the figure.}
\label{rhicspinlayout} 
\end{figure}

A schematic of the RHIC spin complex is shown on the left side of Figure \ref{rhicspinlayout}.  
Each independent RHIC ring,
referred to as the blue and yellow rings, was loaded in this first run with 55 bunches
of polarized protons, at 24 GeV, and then accelerated to 100 GeV.  The bunches were
loaded with different patterns of alternating polarization sign, with three bunches 
with zero polarization, so that the collisions at the experiments included the
combinations ($++$,$+-$,$-+$,$--$,$00$) where the signs refer to (blue,yellow) polarization
sign for the colliding bunches.  

The polarization was measured using separate
polarimeters in the blue and yellow rings, which scattered the polarized protons
from an ultra-thin carbon ribbon which could be inserted into the RHIC beams.  
Scattering in the coulomb-nuclear interference
region is sensitive to the polarization, and the right side 
of Figure \ref{rhicspinlayout} shows the measured asymmetries
from the polarimeters, along with the beam intensities, for a several hour period.
Several RHIC fills are seen in the figure.  When the intensity is seen to rise, the
RHIC ring is being filled with 55 bunches at 24 GeV.  An asymmetry measurement just
after filling is made at injection energy.  During the acceleration to 100 GeV, some
beam is lost and the intensity is seen to reduce.  Measurements of the asymmetry are
then made, typically every 2 hours, on the stored 100 GeV beams.
When we compare the polarization measurements at 24 GeV, before acceleration, and
at 100 GeV, the asymmetry is comparable.  
This is the first use of Siberian Snakes
at high energy, and the polarization was maintained for acceleration through
roughly 150 depolarizing resonances up to 100 GeV, with no adjustment to the Snakes. 
The polarization, using the analyzing power from
24 GeV, was 20\% for this run.

\section{The Physics of the First Run: PHENIX and PP2PP}

The first run, completed this January, 2002, collided transversely
polarized protons at $\sqrt{s}=200$\,GeV, 
$L=10^{30}$\,cm$^{-2}$\,s$^{-1}$, $P=0.2$.  Experiments
measured cross sections and asymmetries for mid-rapidity and forward
pion production, for proton-proton small angle elastic scattering.
Bernd Surrow, in his presentation on the STAR experiment, discusses
forward production.  Here, we will present mid-rapidity and elastic scattering.

\begin{figure}[ht]
\setlength{\unitlength}{\textwidth}
\begin{picture} (0.7,0.35)
\put (0.05,0.0){\mbox{\epsfig{figure=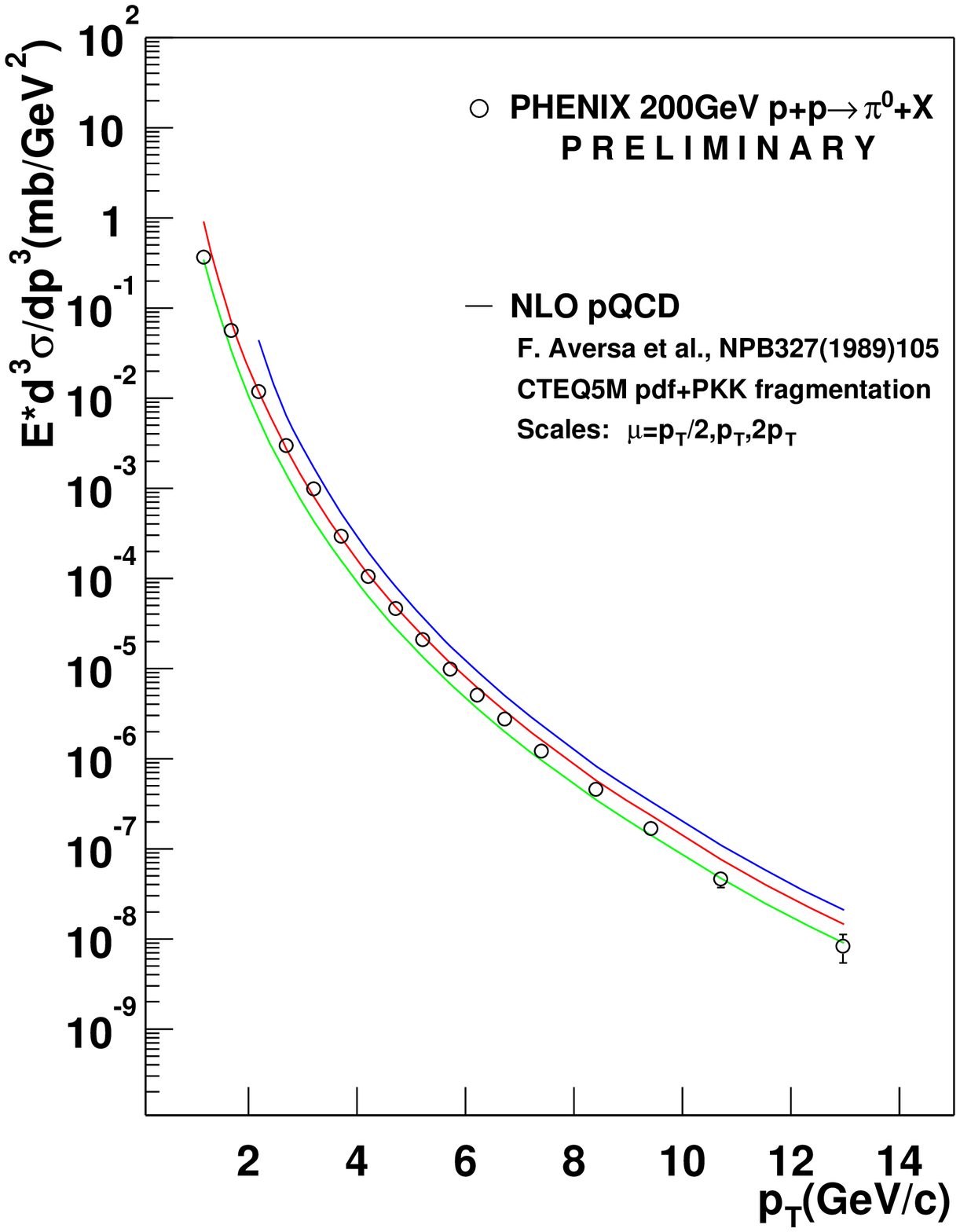,width=5.0cm,clip=}}}
\put (0.60,0.01){\mbox{\epsfig{figure=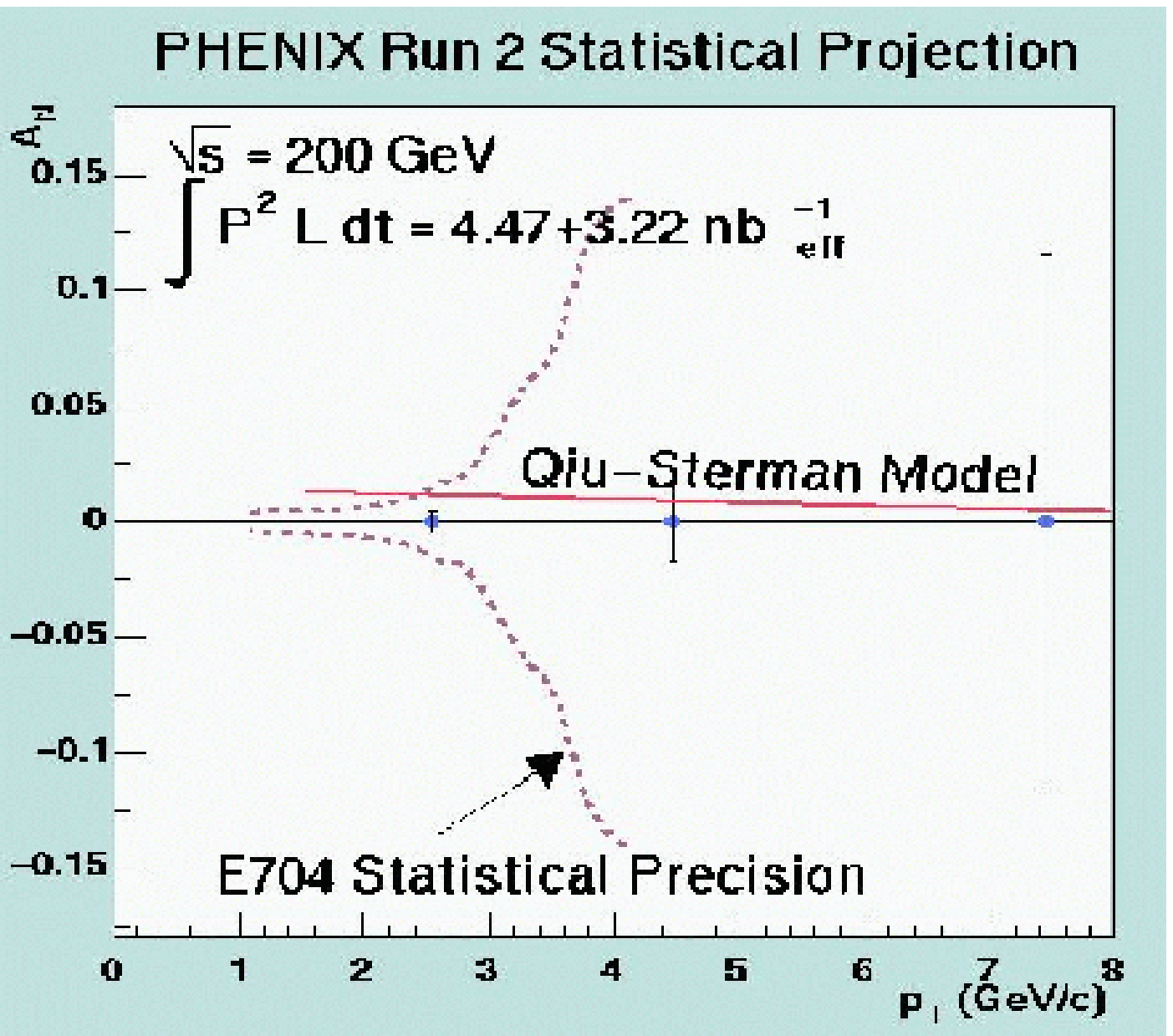,width=5.0cm,clip=}}}
\end{picture}
\caption{\it Preliminary $\pi^0$ cross section vs. $p_T$ at mid-rapidity from
the PHENIX Collaboration, with a published pQCD NLO calculation overlaid (left); 
and the expected sensitivity for the asymmetry $A_N$ for $\pi^0$ (right).  The
dashed lines on the right side indicate the size of the error bars on an earlier
fixed target experiment from Fermilab at much lower energy.}
\label{phenixfig} 
\end{figure}

Measurements of the transverse spin asymmetry $A_N$ probe the transverse
spin structure of the proton.  Measurements were made for $\pi^0$s and charged hadrons
by PHENIX, for colliding transversely polarized protons.
The PHENIX Experiment is one of the two large heavy ion/spin detectors at RHIC \cite{phenix}.
The experiment is a fast detector, featuring excellent photon and electron calorimetry
at mid-rapidity, with tracking and RICH detectors, and muon arms in the forward direction
from each beam.  The electromagnetic calorimeters 
at mid-rapidity were used in the pp run to trigger on energy clusters 
above 1 GeV, and a large sample
of high-$p_T$ $\pi^0$s were collected.  Figure \ref{phenixfig}, left, shows 
the PHENIX Preliminary result for the $\pi^0$ cross section vs. $p_T$\cite{Torii}.  
Also shown on the left figure is a perturbative NLO QCD calculation for the cross
section \cite{aversa}, with three factorization scales.  This is not a fit.  The
agreement of the theory with the data shows that pQCD properly describes the production
of pions at mid-rapidity at RHIC.  This is important to the RHIC spin program, since
it is the interpretation of pp scattering with factorization and pQCD which will
allow us to extract information on the spin structure of the proton at RHIC.
We note also the good agreement down to $p_T$=2~GeV/c, and the relatively small
dependence on the factorization scale.  

The right side of the figure shows the expected
statistical significance of the measurement of the single spin transverse asymmetry
$A_N$.  The figure shows sensitivity at a 1\% level.  An asymmetry observed by 
the HERMES \cite{hermes} and SMC \cite{smc} experiments, interpreted
as transversity, would imply an of order 10\% asymmetry at RHIC \cite{anselmino}.  
Also shown are dashed 
lines giving the
error sizes vs. $p_T$ for a previous mid-rapidity measurement at much lower energy,
E704 at Fermilab \cite{E704}.  The theoretical model shown \cite{qiu_sterman}
is for a higher twist effect, from quark-gluon correlations in the proton.
This mid-rapidity measurement will probe a 
qualitatively new physics region for spin,
in both energy and transverse momentum.

\begin{figure}[ht]
\setlength{\unitlength}{\textwidth}
\begin{picture} (0.7,0.35)
\put (0.05,0.075){\mbox{\epsfig{figure=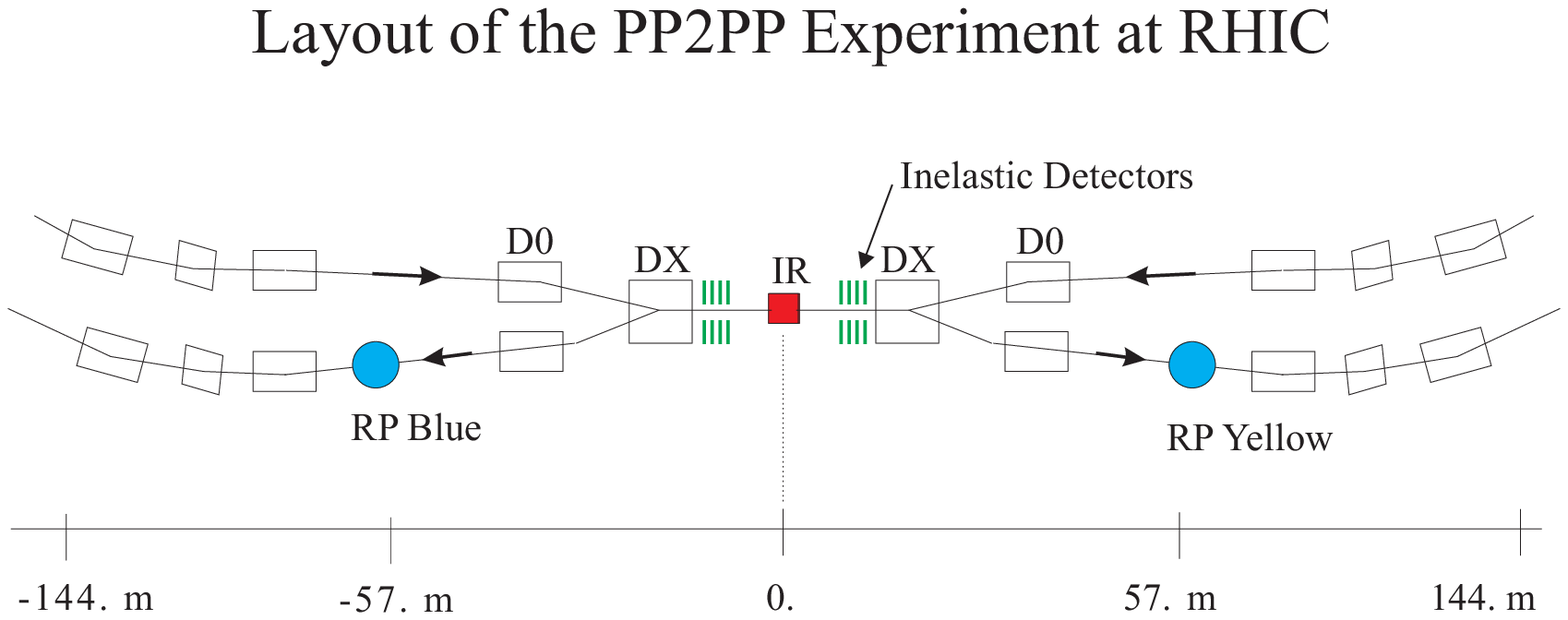,width=7.0cm,clip=}}}
\put (0.6,0.0){\mbox{\epsfig{figure=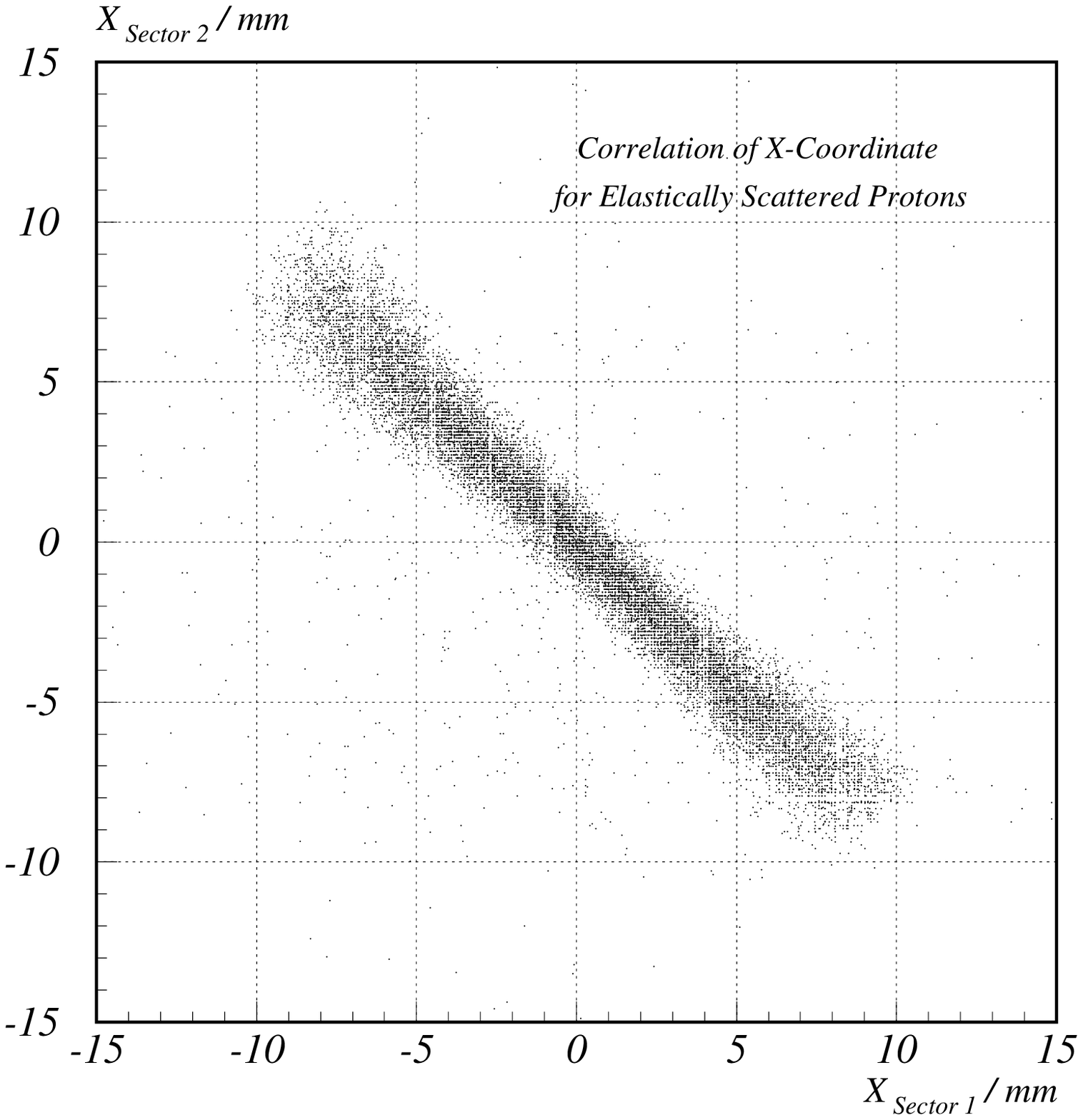,width=6.0cm,clip=}}}
\end{picture}
\caption{\it PP2PP Experiment layout (left) and the observed correlation between the
scattered protons in each beam as seen in silicon detectors near the beams (right).}
\label{pp2ppfig} 
\end{figure}

The PP2PP Experiment \cite{pp2pp} uses devices called Roman Pots to move 
silicon detectors close
to the beams, after the beams are stored in RHIC.  The experiment measures pp elastic
scattering in the coulomb, coulomb-nuclear interference (CNI), and low to mid-t regions.
In this run, the experiment used a 1-day store where the RHIC beams were severely
scraped to reduce their emittance to allow the experiment to make measurements in
the CNI region, for $-t>0.004$\,(GeV/c)$^2$.  These are the first pp measurements for
$\sqrt{s}=200$\,GeV, and the first for spin.  
Figure \ref{pp2ppfig} shows a schematic of the location
of the Roman Pots, left, and on the right side shows the observed correlation
between the scattered proton position in one silicon detector near one beam vs. the 
scattered proton position for the other beam.  This on-line result is remarkably clean,
and the experiment expects to obtain the t-dependence and asymmetry $A_N$ from the
run.

\section{Plan and Conclusions}

The gluon polarization will be measured at RHIC using direct photons, photon plus jet,
and jets.  Anti-quark and quark polarizations will be measured, identified by flavor,
using parity-violating W boson production.  These are the major focus of the RHIC
spin program. 

There are many other processes and probes planned.  For the coming year, when luminosity,
polarization and running time will still be at a premium, jet production should
provide a first look at gluon polarization. 
Other processes are heavy quark production with spin, transversity measurements,
fragmentation processes with spin, searches for new physics using parity violation,
and many others.  The program, described in \cite{bunce}, 
is expected to take 5+ years of RHIC running, which consists of both heavy
ion and polarized proton modes.  This schedule does not include follow-on experiments to pursue
exciting signals and new ideas.

We have important issues to solve to reach our goals: higher polarization and luminosity,
and absolute calibration of the RHIC polarimeters.  The polarization from the AGS was very low
in our first run.  To improve this, we will be using a much faster ramp rate in the AGS next
year.  We are building a fast polarimeter for the AGS based on the RHIC polarimeters
to improve our identification of problem areas.  We are also designing a new strong AGS 
partial Siberian Snake, available for 2005.  For the luminosity, we need to gain a
factor of 100 over this first run.  Many of the anticipated improvement factors are
straight forward, such as a factor 3 improvement using focusing a the interaction
regions with $\beta^*=1$, already used for
heavy ions, and a factor 2 from doubling the number of bunches.  We will see.

In summary, the polarized proton program at RHIC has begun, we have some beautiful data,
and we have a long, exciting road ahead.

\end{document}